\documentclass[prb,twocolumn,showpacs,floatfix]{revtex4}
\usepackage{amssymb}

\usepackage{graphicx}
\usepackage{dcolumn}
\usepackage{bm}

\begin{document}

\title{STM conductance of Kondo impurities on open and structured surfaces}
\author{P. S. Cornaglia}
\author{C. A. Balseiro}

\affiliation{Instituto Balseiro and Centro At\'omico Bariloche, Comisi\'on
Nacional de Energia At\'omica, 8400 San Carlos de Bariloche, Argentina.}

\begin{abstract}
We study the scanning tunneling microscopy response for magnetic atoms on
open and structured surfaces using Wilson's renormalization group. We
observe Fano resonances associated with Kondo resonances and interference
effects. For a magnetic atom in a quantum corral coupled to the confined
surface states, and experimentally relevant parameters, we observe a large
confinement induced effect not present in the experiments. These results
suggest that the Kondo screening is dominated by the bulk electrons rather
than the surface ones.
\end{abstract}

\date{\today}
\pacs{72.15.Qm, 72.10.Fk}
\maketitle

\section{Introduction}

Scanning Tunneling Microscopy\cite{Chen} (STM) has proven to be one of the most useful
tools to measure and characterize low energy electron spectroscopy, at least
on open surfaces where the microscope tip has a direct access to the
physical region under study. In a series of recent works, Kondo impurities 
\cite{Kondo1964, Hewson1993} on the surface of noble metals have been
studied using the STM.\cite{Madhavan1998, Madhavan2001, Ti} In the Kondo problem
there is a characteristic temperature $T_{K}$ that separates the low
temperature from the high temperature regimes. In the low temperature regime
($T<T_{K}$), the spectral density of a Kondo impurity develops a resonance
at the Fermi energy, known as the Kondo resonance, and consequently locally
changes the low energy spectroscopic properties of the system. When the
microscope tip is placed on top of the impurity, the characteristic
conductance vs voltage $V$ shows a typical Fano-like\cite{fano, Plihal2001} structure that is
identified as the fingerprint of the Kondo effect. In this way, the Kondo
behavior of several transition metal impurities on different noble metal
surfaces has been identified. In particular, the temperature dependence of
the Kondo effect has been measured and, by studying Co impurities on
different Cu surfaces, the scaling of the Kondo temperature with the host
electron density has been shown.\cite{Knorr2002}

STM has also been used to create nanostructures on noble metal surfaces by
manipulation of single atoms. This allows to positionate impurities at
special points of the surface or to engineer the environment of the
impurity. In recent experiments the Kondo effect for a $Co$ impurity inside
a quantum corral has been observed.\cite{Manoharan2000}

The Fano like structure of the conductance observed when the STM tip is on
top of a Kondo impurity, can be explained with a simple phenomenological
model which includes an impurity resonant state at the Fermi energy. This
resonance, that represents the Kondo resonance, has a width $k_{B}T_{K}$. In
this model, the differential conductance though the tip $G\left( V\right)
=\partial I/\partial V$ is given by:
\begin{equation}
G(V)=g_{0+}a\frac{q^{2}-1+2xq}{1+x^{2}}.  \label{fano}
\end{equation}
Here $g_{0}$ is a background conductance, $a$ is a constant, $%
x=(eV-\varepsilon_F)/k_{B}T_{K} $ and $q$ is the Fano parameter that depends
on the electronic structure and on the tunneling matrix elements.

For a more quantitative estimation, a many body calculation including the
electron - electron interactions leading to the Kondo resonance is needed.
Moreover, if the impurity is placed on a surface that is structured at a
nanoscopic scale, the local density of states may strongly depend on energy,
an effect that changes the shape of the Kondo resonance \cite{nos_din} and
consequently of the voltage dependence of the STM conductance. A
nanostructure with a characteristic length $L$ has a characteristic energy
scale $\Delta \sim \hbar v_{F}/L$ where $v_{F}$ is the Fermi velocity. In
systems with $\Delta \sim k_{B}T_{K}$ we expect strong effects. Quantum
corrals on the $Cu$ $(111)$ surface can be engineered with the desired
energy scale $\Delta $, however it is not clear how much the surface states
and bulk states contribute to the Kondo effect. Despite of some notorious
effects like the observation, in elliptical corrals, of the so-called mirages,\cite{Manoharan2000} some experiments indicate that the surface states play almost no role in the
development of the Kondo effect.\cite{Knorr2002} If so, the Kondo screening would not be
affected by the corral that confines only surface states. A detailed
analysis of the STM conductance can be used to elucidate this point, however
to do that the surface nanostructure and the Kondo correlations have to be
treated on the same footing.

In what follows we present results for the tip conductance on top of a Kondo
impurity. Using the numerical renormalization group, we evaluate the finite
temperature conductance for an impurity on open and clean surfaces and on structured ones.

\section{The Model and Results}

In this section we present the model and briefly discuss how the STM
conductance is calculated using Wilson's renormalization group. We then
present the results for different situations. Our starting point is an
extended Anderson model for magnetic impurities described with the following
Hamiltonian:\cite{Hewson1993}

\begin{eqnarray}
H_{AM} &=&\sum_{\sigma }\varepsilon _{d}d_{\sigma }^{\dagger }d_{\sigma
}+Ud_{\uparrow }^{\dagger }d_{\uparrow }d_{\downarrow }^{\dagger
}d_{\downarrow }+\sum_{k,\sigma }\varepsilon _{k}c_{k\sigma }^{\dagger
}c_{k\sigma }  \nonumber \\
&&+\sum_{k,\sigma }(V_{k}^{*}c_{k\sigma }^{\dagger }d_{\sigma
}+V_{k}d_{\sigma }^{\dagger }c_{k\sigma })  \nonumber  \label{ham} \\
&&+\sum_{k k^\prime \sigma \sigma^\prime}U^{k k^\prime}_{dc}d_{\sigma
}^{\dagger }d_{\sigma } c_{k\sigma^\prime }^{\dagger }c_{k^{\prime
}\sigma^\prime }
\end{eqnarray}
where the operator $d_{\sigma }^{\dagger }$ creates an electron with spin $%
\sigma $ at the impurity orbital with energy $\varepsilon _{d}$ and Coulomb
repulsion $U$, $c_{k\sigma }^{\dagger }$ creates an electron in an extended
state with quantum numbers $k$ and $\sigma $ and energy $\varepsilon _{k}$.
The last term is a Falicov-Kimbal like term which takes into account a Coulomb repulsion between the electrons at the impurity orbital and the ones at extended
states. The quantum number $k$ includes the band index and crystal momentum.
In the case of impurities on the Cu (111) surface there is a surface band
that is relevant for determining the STM conductance.\cite{Knorr2002}

The STM conductance at low temperatures is given by\cite{Mahan}
\begin{equation}\label{GV0}
G(V)=\frac{4\pi e^{2}}{\hbar }\rho _{t}\rho (\varepsilon _{F}+eV),
\end{equation}
where $eV$ is the voltage drop from the tip to the sample, and $\rho _{t}$
is the tip density of states that is assumed to be constant around the Fermi
energy. In the linear response regime, the quantity $\rho (\varepsilon
_{F}+eV)$ is given by 
\begin{equation}
\rho (\varepsilon _{F}+eV)=-\frac{1}{\pi }Im\mathcal{G}(\varepsilon _{F}+eV).
\label{green}
\end{equation}
Here $\mathcal{G}(\omega )$ is the retarded Green function of the operator $%
t_{c}\psi _{\sigma }^{\dagger }(R)+t_{d}d_{\sigma }^{\dagger }$, $t_{c}$ and 
$t_{d}$ are matrix elements for tunneling of an electron from the tip to the
conduction band states and impurity orbital respectively and $\psi _{\sigma
}^{\dagger }(R)$ is the field operator that creates an electron on the
surface at coordinate $R$. In what follows we consider that the tip is on
top of the impurity and take $R\equiv R_{i\text{ }}$to be the impurity
coordinate.

The Green function is calculated using the numerical renormalization group
(NRG). For a general description of the method we refer to Refs. %
\onlinecite{Wilson1975} and \onlinecite{Costi1994}, here we give only some
details that are specific to our calculation. The starting point of the NRG
is to reduce the Hamiltonian (\ref{ham}) to that of a linear chain

\begin{eqnarray}
H_{AM} &=&\sum_{\sigma }\varepsilon _{d}d_{\sigma }^{\dagger }d_{\sigma
}+Ud_{\uparrow }^{\dagger }d_{\uparrow }d_{\downarrow }^{\dagger
}d_{\downarrow }+\sum_{\sigma }W_{0}f_{0\sigma }^{\dagger }f_{0\sigma }
\label{rgham} \nonumber\\
&&+V\sum_{\sigma }(f_{0\sigma }^{\dagger }d_{\sigma }+d_{\sigma }^{\dagger
}f_{0\sigma })+U_{dc}\sum_{\sigma \sigma ^{\prime }}d_{\sigma }^{\dagger
}d_{\sigma }f_{0\sigma ^{\prime }}^{\dagger }f_{0\sigma ^{\prime }} 
\nonumber \\
&&+\sum_{\sigma ,n=0}\lambda _{n}(f_{n\sigma }^{\dagger }f_{n+1\sigma
}+f_{n+1\sigma }^{\dagger }f_{n\sigma })
\end{eqnarray}
The states described by the operators $f_{n\sigma }^{\dagger }$ are Wilson's
orbitals centered at the impurity with the state $|0,\sigma \rangle $
associated with $f_{0\sigma }^{\dagger }$ being a Wannier like orbital
centered at the impurity coordinate. The parameters $\lambda _{n}$ generate
a logarithmic discretization of the conduction band, and $W_{0}$ describes
the impurity potential scattering. The Hamiltonian (\ref{rgham}) is
diagonalized numerically using an iterative controlled scheme. At each
iteration step $N$, a truncated Hamiltonian with $N$ orbitals is used and
the many body states within an energy scale $\omega _{N}$ are evaluated. The
thermodynamic and dynamical properties at each energy scale can then be
calculated recursively. 

As it is usually done to describe the Kondo problem, Hamiltonian (\ref{rgham}%
) includes a single channel that is due to all the conduction bands. In the
case of the $Cu(111)$ surface there is a surface band and bulk states that
contribute to the density of states at the Fermi energy. Wilson's orbitals
are assumed to be a linear combination of surface and bulk states, in
particular the Wannier state $|0,\sigma \rangle $ contains a surface and a
bulk component determined by the respective hybridization matrix elements $%
V_{k}$.

To calculate the STM conductance when the tip is on top of the impurity, we
take the field operator $\psi _{\sigma }^{\dagger }(R_{i})$ to be $%
f_{0\sigma }^{\dagger }$ and calculate the retarded Green function of the
operator $t_{c}^{\prime }f_{0\sigma }^{\dagger }+t_{d}d_{\sigma }^{\dagger }.
$ The matrix element $t_{c}^{\prime }$ depends on how much the Wannier
orbital $|0,\sigma \rangle $ contains of the surface or bulk states. To be
more specific on the nature of the Wannier orbital $|0,\sigma \rangle $ and
on the matrix element $t_{c}^{\prime }$, a detailed model of the electronic
structure of the metal host and of the electron transfer from the tip to the
surface is required. 

The retarded Green function of Eq. (\ref{green}) is calculated as in Ref. %
\onlinecite{Costi1994} and

\begin{eqnarray}
\rho (\varepsilon _{F}+eV) &=&-\frac{1}{\pi }[\mathcal{G}(\omega +i0^{+})-%
\mathcal{G}(\omega -i0^{+})]_{\omega =\varepsilon _{F}+eV}  \nonumber \\
&=&\sum_{\lambda ,\lambda ^{\prime }}|M_{\lambda ,\lambda ^{\prime }}|^{2}%
\frac{(e^{-E_{\lambda }/k_{B}T}+e^{-E_{\lambda ^{\prime }}/k_{B}T})}{Z(T)}%
\times  \nonumber \\
&&\delta (\varepsilon _{F}+eV-(E_{\lambda ^{\prime }}-E_{\lambda }))
\label{ro}
\end{eqnarray}
where $Z(T)$ is the grand partition function 
\begin{equation}
Z(T)=\sum_{\lambda }e^{-E_{\lambda }/k_{B}T}
\end{equation}
and $M_{\lambda ,\lambda ^{\prime }}=\langle \lambda |t_{c}^{\prime
}f_{0\sigma }^{\dagger }+t_{d}d_{\sigma }^{\dagger }|\lambda ^{\prime
}\rangle $. At each energy scale $\omega _{N}$ the discrete spectra is
smoothed by replacing the delta functions by a smooth distribution. In what
follows we use a logarithmic Gaussian that has been proven to give the best
results.

\subsection{Impurity on a clean surface}

\begin{figure}[tbp]
\includegraphics[width=7.5cm,clip=true]{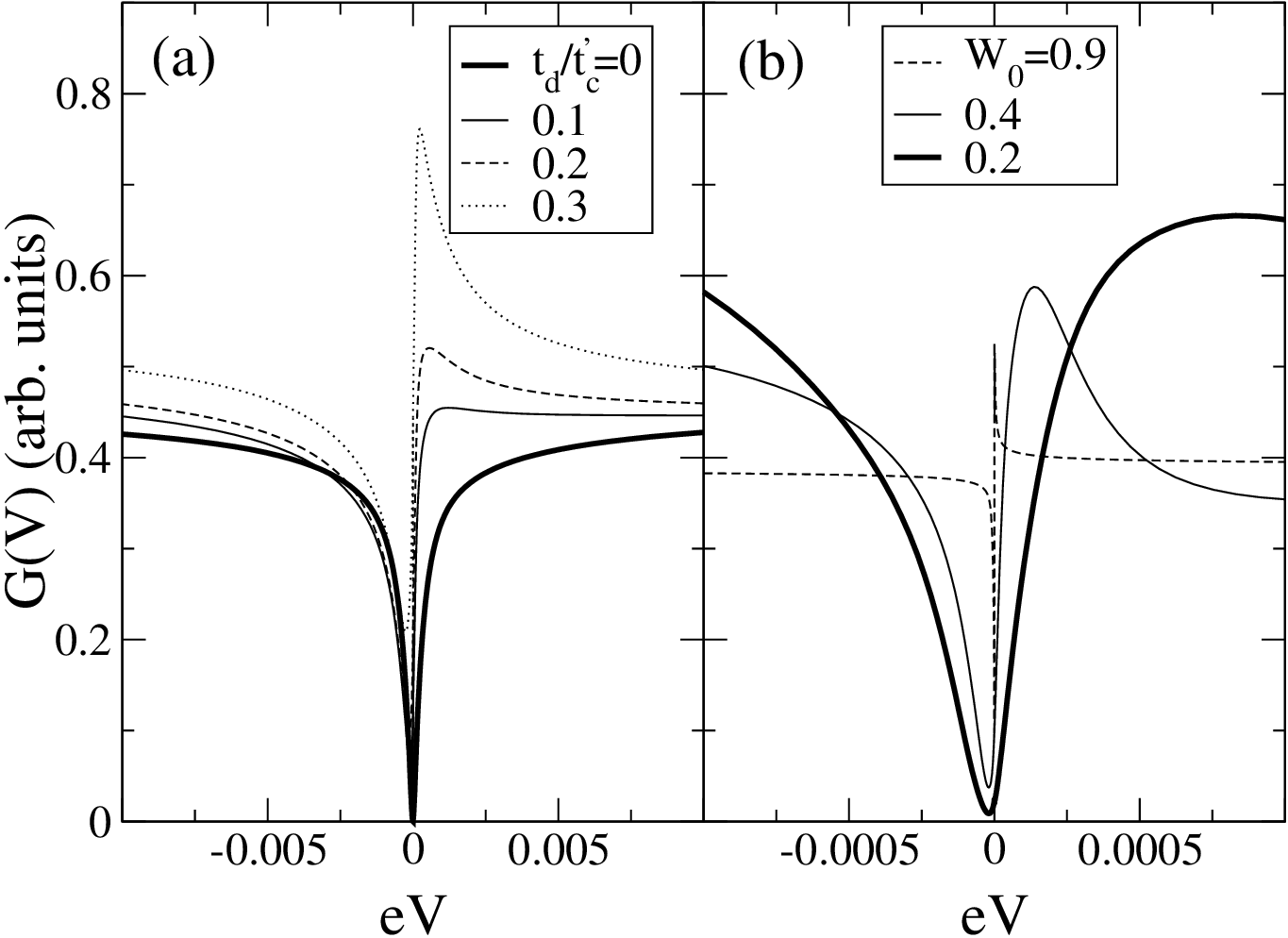}
\caption{STM conductance vs voltage, parameters in units of half the
bandwidth are $\varepsilon _{d}=-U/2=-0.5$, $U_{ds}=0$, $V=0.2$ and (a) $W_{0}=0$ and different values of $t_{d}/t_{c}^\prime$ as indicated in the box; (b) $t_{d}=0$
and $W_{0}\neq 0$. In both (a) and (b) a Fano lineshape is obtained. }
\label{GV}
\end{figure}

\begin{figure}[tbp]
\includegraphics[width=7.5cm,clip=true]{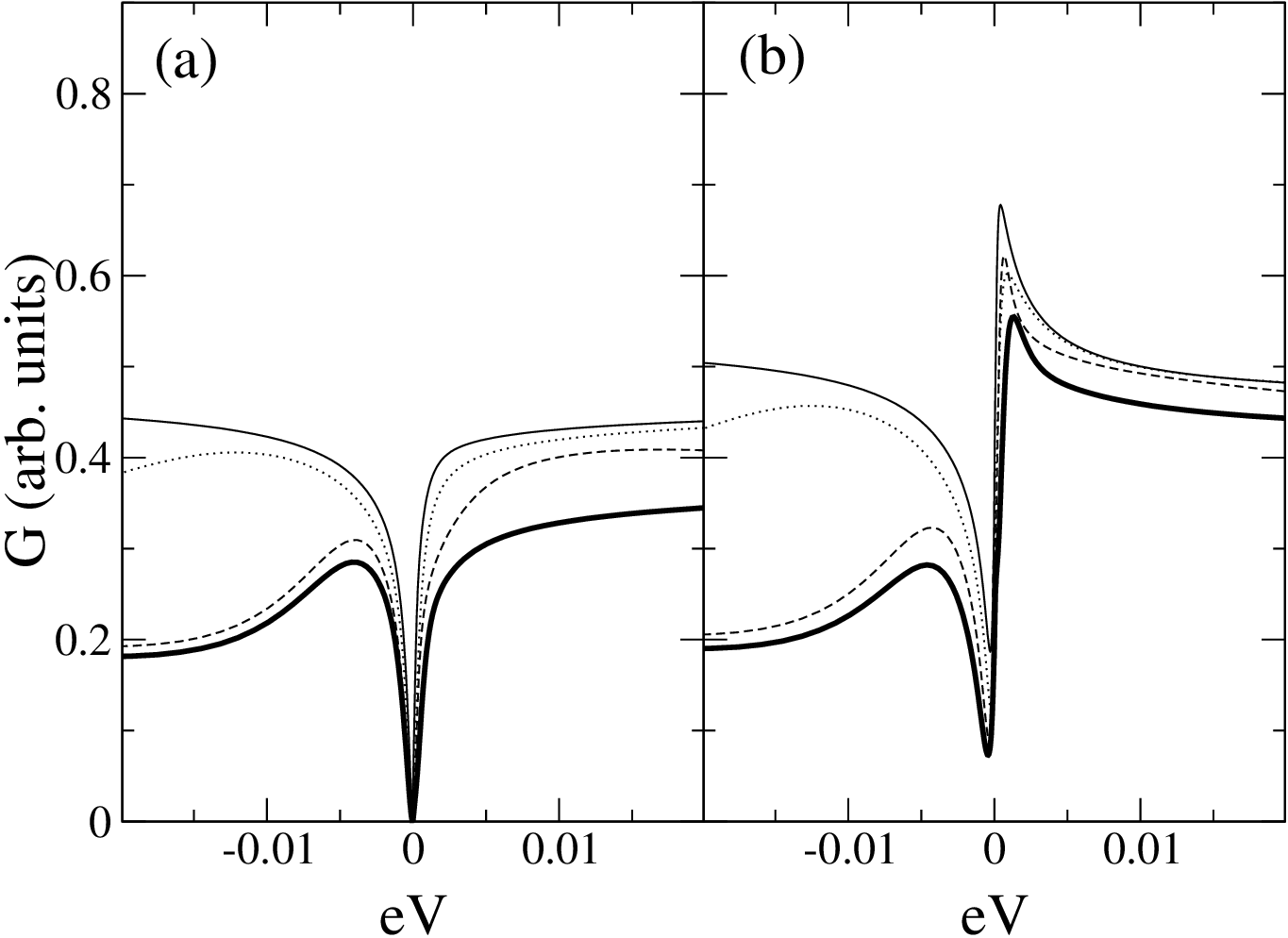}
\caption{Linear conductance spectra for different values of $%
U_{dc}=0.05$ (thin solid line), $0.1$ (dotted line), $0.15$ (dashed line), $0.2$ (solid thick line)  and $\varepsilon_d=-U/2$. (a) $t_{d}/t_{c}^{\prime }=0.1$ (b) $t_{d}/t_{c}^{\prime }=0.3$. Other parameters as in Fig. 2.}
\label{GVU}
\end{figure}

\begin{figure}[tbp]
\includegraphics[width=7.5cm,clip=true]{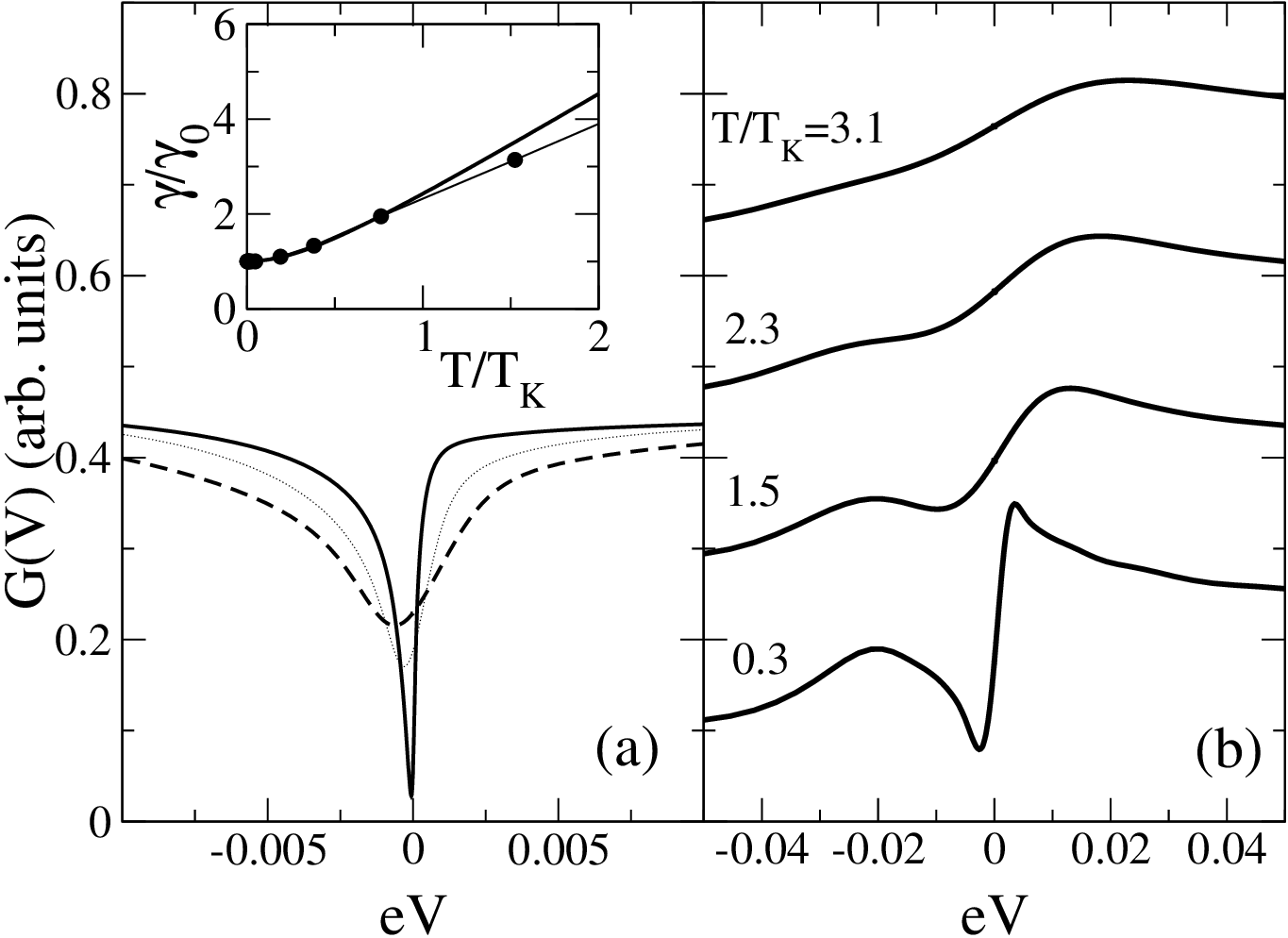}
\caption{(a) Linear conductance spectra for $t_{d}/t_{c}^{\prime }=0.1$ and
different temperatures: $T/T_{K}=0$ (solid line), $0.75$ (dotted line), and $1.5$ (dashed line). Inset: Fermi liquid theory fit to the width $\gamma $
of the Fano resonance as a function of temperature. (b) Linear conductance
spectra in an asymmetric situation $\varepsilon _{d}=-0.2$, $U=1$ and $%
t_{d}/t_{c}^{\prime }=0.35$ for different temperatures as indicated (the
spectra at different temperatures have been shifted vertically for clarity).}
\label{GVT}
\end{figure}

Here we present the results for a Kondo impurity on an open surface. In an
energy scale of several $k_{B}T_{K\text{ , }}$the bare density of conduction
states is constant around the Fermi energy. Then the calculation can be done
with the usual band discretization first proposed by Wilson. The results
obtained for the conductance are shown in Figs. \ref{GV}, \ref{GVU} and \ref
{GVT}.

The STM conductance depends on the following parameters: $k_{B}T_{K}$ that
gives the low energy scale where the local spectroscopic properties are
dominated by the impurity, the temperature $k_{B}T$ and the Fano parameter $q
$ that combines the tunneling matrix elements and some aspects of the band
structure. If the Kondo resonance is represented by a simple resonant state,
the Fano parameter is given by 
\begin{equation}
q=\frac{t_{d}+t_{c}^{\prime }Re\sum (0)}{-t_{c}^{\prime }Im\Sigma(0)}\label{q}
\end{equation}
where $\Sigma (\omega )=\sum_k V_{k}/(\omega -\varepsilon _{k})$ is an effective
band propagator.\cite{Zawadowski} For a system with electron-hole symmetry the real part of $\Sigma (\omega )$ at the Fermi energy is zero and the parameter $q$ is proportional to the matrix element $t_{d}$ that describes the direct
tunneling into the localized state. In this case, $q$ accounts for the
interference between the two tunneling channels. In general, real systems
have a non zero $Re\Sigma (0)$ and there is no need to assume a direct
tunneling to the localized state to have an asymmetric ($q\neq 0$) Fano
resonance. In the NRG approach, the use of a symmetric conduction band
simplifies the calculation and most of the numerical work on the Kondo
effect has been done with this symmetry. We have introduced the potential
scattering term $W_{0}$ to break the electron-hole symmetry of the band. The
Fano resonance then depends on the conduction band parameters and in
agreement with Ref. \onlinecite{Zawadowski}, with a non zero $W_{0}$ we
obtain asymmetric line shapes even for $t_{d}=0$. 

In Fig. \ref{GV} the conductance vs applied voltage is shown for $%
\varepsilon _{d}=-U/2$ and $U_{dc}=0$, all parameters are given in units
of half the bandwidth $D=1$. In Fig. \ref{GV}(a) the results for $%
W_{0}=0$ and a set of the ratios $t_{d}/t_{c}^{\prime }$ are shown, this ratio determines the structure of the conductance. In Fig. \ref{GV}(b), by independently varying $W_{0}$ for $t_{d}=0$ we show that in a full NRG many body calculation, the low energy
STM conductance is still given by a single parameter $q$ as given by
equation (\ref{q}). As $W_0$ is increased the resonance narrows indicating a lowering of the Kondo temperature. This is due to a decrease in the local density of states at the Fermi energy. The increase of $W_0$ also induces an increase of the Fano parameter. For simplicity, in what follows we take symmetric band ($Re\Sigma(0)=0$) where the Fano parameter is given by $q=t_{d}/t_{c}^{\prime }\pi\rho V$, where $\rho$ is the density of states of the band.

Our calculation, that up to now mainly reproduces known results, shows that
the NRG gives reliable results when used to evaluate the STM conductance. In
what follows we present some results that include the behavior of the
conductance when the energy of the localized orbital approaches the Fermi
energy, the effect of the Falicov-Kimbal term $U_{dc}$ and the temperature
dependence of the conductance.

\begin{figure}[tbp]
\includegraphics[width=7.5cm,clip=true]{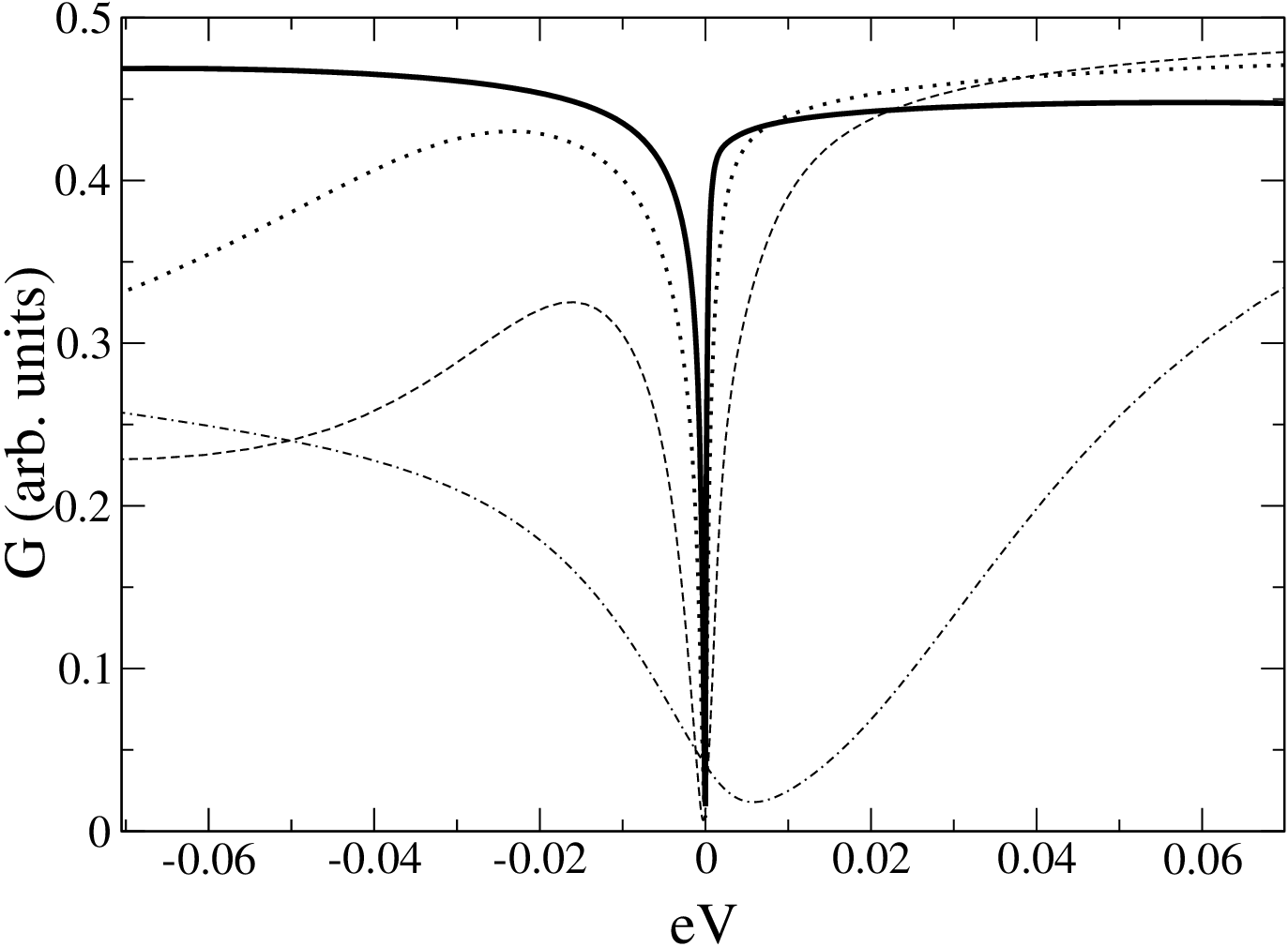}
\caption{Linear conductance spectra for different values of the impurity 
level energy $\varepsilon _{d}=-0.5U,\,-0.3U,\,-0.2U$, and $-0.05U$. As $%
\varepsilon _{d}$ approaches the Fermi level, the Fano resonance becomes
wider. Other parameters are $U=1$, $t_{d}/t_{c}^\prime=0.1$, $V=0.2$, and $W_{0}=U_{ds}=0$.
}
\label{GVE}
\end{figure}

We start by discussing the conductance dependence on the localized orbital
energy $\varepsilon _{d}$. As $\varepsilon _{d}$ is increased and approaches
the Fermi energy, first the Kondo temperature increases and then the
impurity enters an intermediate valence regime. In the conductance shown in
Fig. \ref{GVE} these effects are clearly observed. A notorious effect for
intermediate values of $\varepsilon _{d}$ is the occurrence of an asymmetry
for finite voltages: away from the Fermi level the values of $G(V)$ are
smaller for $V<0$ than for $V>0$. This effect is due to the localized
orbital resonance that is close to the Fermi energy. As $\varepsilon _{d}$
increases even more (not shown) the resonance crosses the Fermi level, the impurity
occupation is reduced and the conductance shows a broad structure with a
width that is given by the resonance width $\Gamma $ rather than by the
Kondo temperature. As we now show, the effect of a non zero
electron-electron interaction $U_{dc}$ is equivalent to a shift in the
impurity level. The parameter $U_{dc}$ is irrelevant, that means that in the
Kondo regime the zero temperature fixed point does not depend on this
parameter and the low temperature thermodynamic behavior is universal as for 
$U_{dc}=0$. However this interaction may change some details in the low
energy spectral structure for energies of the order or larger than $%
k_{B}T_{K}$. We have analyzed the effect of $U_{dc}$ on the STM conductance
and the results are summarized in Fig. \ref{GVU}. As $U_{dc}$ increases the
Fano line shape changes and two different effects can be observed: \textit{i}%
) the width of the conductance minimum increases indicating an increase in
the Kondo temperature and \textit{ii}) the voltage dependence of the
conductance becomes more asymmetric. These two points are consistent with an
effective shift the localized level toward the Fermi energy. 
The asymmetry in $G(V)$ obtained when the resonant level approaches the
Fermi level has not been discussed in detail in the literature, however
experimental results for $Co$ on $Cu(111)$ and for $Ti$ on $Ag$ show this
type of effect. In Ref. \onlinecite{Ti} the raw experimental has a 
marked asymmetriy in the conductance that was attributed to a broad
resonance lying just above the Fermi energy. Our results may explain the
asymmetry as due to the same localized orbital that generates the Kondo
resonance.

Now we present finite temperature results and analyze the temperature
dependence of the conductance. To do this we calculate the finite temperature spectral densities following Refs. \onlinecite{nos_din} and \onlinecite{Costi1994} and use the finite temperature extension of Eq. (\ref{GV0})
\begin{equation}
G(V,T)=-\frac{4\pi e^{2}}{\hbar }\rho _{t}\int d\varepsilon \frac{\partial f(\varepsilon - eV)}{d\varepsilon}\rho (\varepsilon),
\end{equation}
where $f(\varepsilon)$ is the Fermi function.
In Fig. \ref{GVT}(a) the conductance for a
small $q$ and $W_{0}=U_{dc}=0$ is shown for different temperatures. As the
temperature increases the width of the low voltage structure increases. In
agreement with the experimental results of Ref. \onlinecite{Ti},
our results show a width that increases according to the low temperature
Fermi liquid theory. In the inset of Fig. \ref{GVT} (a) the numerical width is
shown for different temperatures, the thick continuous line is a fit with the
Fermi liquid expression $\gamma =\frac{\gamma_0}{\sqrt{2}k_BT_K}\sqrt{(\pi k_{B}T)^{2}+2(k_{B}T_{K})^{2}}$%
. In Fig. \ref{GVT}(b) the temperature dependence of a system with $%
\varepsilon _{d}=-0.2$ is shown. There is remarkable resemblance between
some of these curves and the experimental results of $Ti$ on $Ag$. 

\subsection{Impurity on a structured surface}

We now present results for the STM conductance when the Kondo impurity is on
a structured surface. The main effect of a nanostructure on the surface is
to produce an energy dependence of the local density of states on a scale $%
\Delta $ that may be comparable with $k_{B}T_{K}$. If this happens the Kondo
effect may be very much affected as discussed in previous works. Possible
realizations of this situation may be obtained by placing impurities in
quantum corrals, on small islands or in any structure with a characteristic
size of the order of the hundred \AA. The problem then becomes much more
complicated mainly because of two reasons: on the one hand, the
nanostructure of the host and the Kondo correlations have to be treated on
the same footing. On the other hand, a detailed model for the metallic
surface and the effect of the nanostructure on the bulk and surface states
is needed. In general, in surfaces like the $Cu(111)$ where there is a
surface band, we expect structures like the quantum corrals or small islands
to confine surface states keeping the bulk states nearly unaffected. If so,
these structures would change the Kondo effect according to the
participation of the surface states on the magnetic screening of the
impurity. This point is still controversial: while the observation of
quantum mirages seems to indicate that surface modes actively participate in
the Kondo screening, other experimental results suggest that the Kondo
effect is due to bulk states. The STM conductance could be used to clarify
this point.

Due to the lack of a detailed model suitable to be treated with the
renormalization group approach, we resort to a simple and phenomenological
model to study the effect of structured local densities of states. The model
reduces the problem to a single effective conduction band. As before, we
assume that Wilson's orbitals are built as linear combination of the surface
or bulk states. The surface nanostructure affects mostly the surface
component of these states. In order to describe a local density of states
with a structure in the desired energy range, we resort to the method
described in Refs. \onlinecite{nos_din} and \onlinecite{nos_termo}. The
calculation includes a potential barrier that confines the electrons. The
local density of states so obtained in the absence of impurity has the form
of resonant states. This type of resonances have been observed using the STM
in atomic islands or terraces on the $Cu(111)$ surfaces.\cite{Veuillen} If surface states
participate actively on the Kondo screening, the modulation of the relevant
local density of states would be large indicating that the active electrons
are confined by the nanostructure (a large potential barrier). Conversely,
if the Kondo effect is mostly due to the bulk states, the local density of
states modulation would be very week. In this case we use a small effective
barrier. The model is meant to make only qualitative estimations,
nevertheless as we show below some important conclusions can be drawn from
our analysis. 
\begin{figure}[tbp]
\includegraphics[width=7.5cm,clip=true]{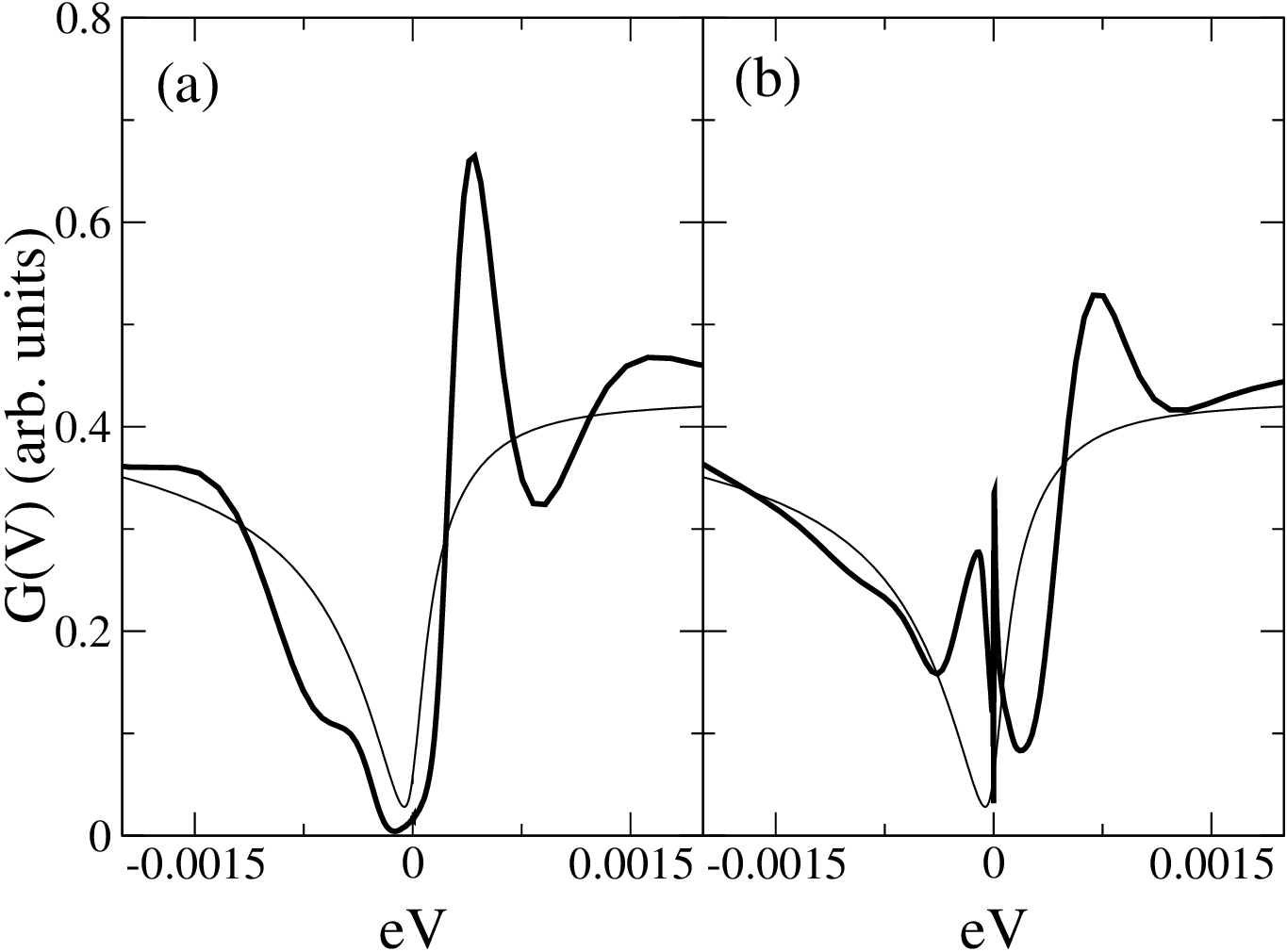}
\caption{Linear conductance spectra for an impurity in a confined system $%
T_{K}\sim \Delta$ and $\delta/\Delta\sim0.35$ (a) at-resonance (b) off-resonance. Other parameter are $\varepsilon_d=-U/2$, $V=0.2$, $t_d/t_c^\prime=0.1$, and $W_0=U_{dc}=0$. For comparison the lineshape for an open surface is shown ( thin line).}
\label{GVB}
\end{figure}

As stated above, a nanostructure introduces a new length scale and the
corresponding energy scale $\Delta .$ Since the number of microscopic
parameters of the model is large, we just distinguish to cases: large
structures with $\Delta \lesssim k_{B}T_{K}$ and small nanostructures with $%
\Delta >k_{B}T_{K}$. Another important aspect of the Kondo effect in
nanostructures concerns the position of the Fermi level relative to the
structure of the local density of states. In this respect we consider the
two extreme cases: the Fermi energy lying at a resonance (at-resonance case)
or between two resonances (off-resonance case). The quantum corrals on $%
Cu(111)$ with characteristic sizes of the order of $100\text{\AA}$ correspond to the small nanostructure category. The elliptical corrals of Ref. %
\onlinecite{Manoharan2000} have the Fermi energy at a resonance.

Figure \ref{GVB} illustrates the case of a relatively large nanostructure ($%
\Delta \sim k_{B}T_{K}$) for the at-resonance and off-resonance cases. The
calculation was done with a small $q$ and an intermediate barrier. The low
temperature STM conductance shows a Fano like structure modulated by a
superstructure of peaks separated approximately by the characteristic energy 
$\Delta $. For the at-resonance case the central valley slightly broadens
while for the off-resonance case narrow structures at low voltages are
obtained. As the barrier height decreases the amplitude of the oscillations
in the host local density of states decreases and the conductance continuously approaches the open surface result.

\begin{figure}[tbp]
\includegraphics[width=7.5cm,clip=true]{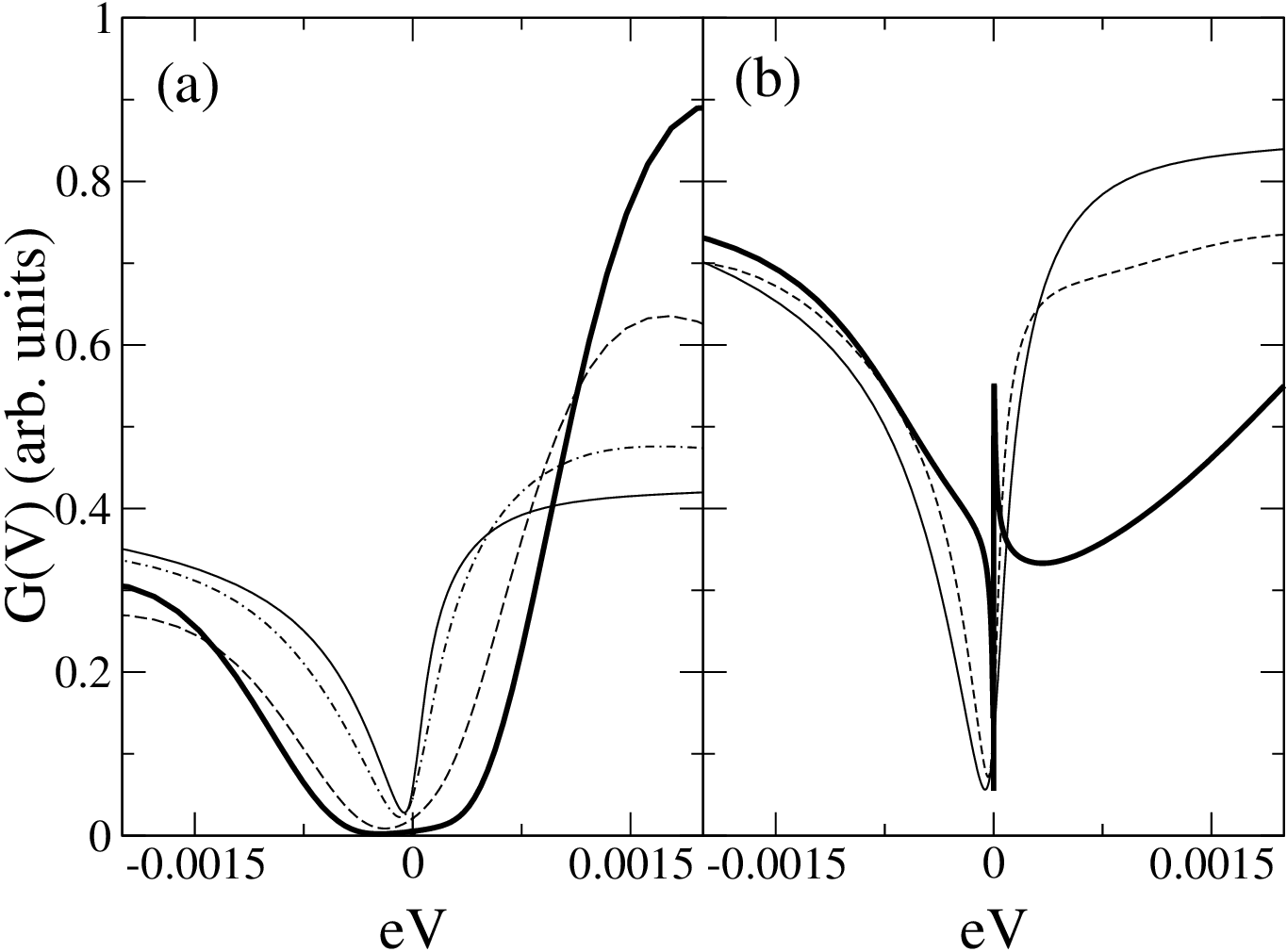}
\caption{Linear conductance spectra for an impurity in a confined system with $t_d/t_c^\prime=0.1$, $T_{K}\sim \Delta/10$ and different widths $\delta$ of the resonant states (a) at-resonance without barrier (thin solid line) and $\delta/\Delta\sim 0.63$, $0.42$, $0.21$ (thick solid line) (b) off-resonance without barrier (thin solid line) and $\delta/\Delta\sim 0.63$, $0.42$ (thick solid line).}
\label{GVB1}
\end{figure}

Figure \ref{GVB1} shows the results for a small system ($\Delta >k_{B}T_{K}$%
). In this case the superstructure is not visible since $\Delta $ is out of
scale. For the at-resonance (Fig. \ref{GVB1}(b)) case and large barriers, the Kondo temperature
aincreases and the Fano structure is broadened. This is due to an increase of
the local density of states at the Fermi energy $\varepsilon _{F}$. The
Kondo temperature has an exponential dependence with the density of states
and consequently the broadening is very important for intermediate or large
barriers. For the off-resonance case (Fig. \ref{GVB1}(b)), the density of
states decreases and the structure narrows, for large barrier a narrow Fano
structure with a large $q$ mounted on a background with a pronounced minimum
at $\varepsilon _{F}$ is obtained.

\section{Summary and Discussion}

We have analyzed the STM conductance $G(V)$ for Kondo impurities on clean
and structured surfaces. Starting with an Anderson model we used the NRG to
calculate the physical properties and voltage dependence of the conductance.
For clean surfaces we showed that the numerical approach gives reliable
results and reproduces the Fano type STM response. In particular, we have
studied in some detail how $G(V)$ depends on different microscopic
parameters and on temperature. As the impurity level is shifted away from
the symmetric case $\varepsilon _{d}=-U/2$, either by changing its energy or
by means of a Falicov-Kimbal interaction, the Fano line is distorted. For a set of
physical parameters, both the shape and the temperature dependence of the
conductance are in good agreement with experimental observations. The
temperature dependence is also in good agreement with the low temperature
Fermi liquid theory.

To discuss the STM conductance of impurities in structured surfaces we used
an effective one channel model. In surfaces like the $Cu(111)$ one, where
there is a surface band, this effective channel represents a linear
combination of surface and bulk sates. Since a surface structure perturbs
much more the surface modes than the bulk modes, the sensitivity of the
Kondo effect to the structure gives an idea of the surface mode
participation in the Kondo screening. We have shown that small structures
like the quantum corrals, that are quite efficient in confining surface
electrons, should produce important changes in the Kondo temperature and on
the STM response if the surface modes were active in the Kondo physics. The
experiments show that the Fano line has essentially the same width when the
impurity is on a clean surface or inside a corral, only small changes in the
parameter $q$ are observed.\cite{mano} This fact is a clear indication that
the surface modes can sense the Kondo effect but are not important in
determining it. This conclusion is in agreement with estimation of the
microscopic parameters for surface and bulk states.\cite{ana}

This work was partially supported by the CONICET and ANPCYT, grants N. 02151
and 99 3-6343.


\begin{thebibliography}{99}

\bibitem{Chen} C. J. Chen, {\it Intorduction to Scanning Tunneling Microscopy} (Springer, Berlin, 1996).
\bibitem{Kondo1964}  J. Kondo, Prog. Theor. Phys. \textbf{32}, 37 (1964).

\bibitem{Hewson1993}  A. C. Hewson, \textit{The Kondo problem to heavy
fermions} (Cambridge University Press, Cambridge, England, 1993).

\bibitem{Madhavan1998}  V. Madhavan, W. Chen, T. Jamneala, M. F. Crommie,
and N. S. Wingreen, Science \textbf{280}, 567 (1998).

\bibitem{Ti}  K. Nagaoka, T. Jamneala, M. Grobis, and M. F.
Crommie, Phys. Rev. Lett. \textbf{88}, 077205 (2002).

\bibitem{Madhavan2001} V. Madhavan, W. Chen, T. Jamneala, M. F. Crommie,
and N. S. Wingreen, Phys. Rev. B \textbf{64}, 165412 (2001). 

\bibitem{fano} U. Fano, Phys. Rev.  {\bf 124}, 1866 (1961).

\bibitem{Plihal2001} M. Plihal and J. W. Gadzuk, Phys. Rev. B {\bf 63}, 085404 (2001).

\bibitem{Knorr2002}  N. Knorr, M. A. Schneider, L. Diekh\"{o}ner, P. Wahl,
and K. Kern, Phys. Rev. Lett. \textbf{88} 096804 (2002).

\bibitem{Manoharan2000}  H. C. Manoharan, C. P. Lutz, and D. M. Eigler,
Nature (London) \textbf{403} 512 (2000).


\bibitem{nos_din}  P. S. Cornaglia and C. A. Balseiro, Phys. Rev. B \textbf{%
66}, 174404 (2002).

\bibitem{Mahan} G.D. Mahan, \textit{Many Particle Physics} (Plenum Press, New York, second edition, 1993).

\bibitem{Wilson1975}  K.G. Wilson, Rev. Mod. Phys \textbf{47}, 773 (1975);
H.R. Krishna-murthy, J.W. Wilkins and K.G. Wilson, Phys. Rev. B \textbf{21},
1044 (1980).

\bibitem{Costi1994}  T. A. Costi, A. C. Hewson and V. Zlatic, J. Phys. Cond.
Matt. \textbf{6}, 2519 (1994).

\bibitem{Zawadowski}  O. \'{U}js\'{a}ghy, J. Kroha, L. Szunyogh, and A.
Zawadowski, Phys. Rev. Lett \textbf{85}, 2557 (2000).

\bibitem{nos_termo}  P. S. Cornaglia and C. A. Balseiro, Phys. Rev. B 
\textbf{66}, 115303 (2002).

\bibitem{Veuillen} S. Pons, P. Mallet, and J. Y. Veuillen, Phys. Rev. B {\bf 64}, 193408 (2001).

\bibitem{mano}  H. C. Manoharan, private communication.

\bibitem{ana}  A. M. Llois, private communication.
\end{thebibliography}
\end{document}